\pdfoutput=1
\documentclass[english]{revtex4-1}
\usepackage[latin9]{inputenc}
\usepackage{verbatim}
\usepackage{graphicx}
\usepackage{babel}
\begin{document}

\title{Measuring Charged Particle Multiplicity with Early ATLAS Public Data}

\author{Gözde Üstün$^{\dagger}$, Erol Barut$^{\ddagger}$, Erhan Bekta\c{s}$^{\ddagger}$,
V. Erkcan Özcan$^{\ddagger}$}

\affiliation{$^{\dagger}$Y\i ld\i z Technical University, $^{\ddagger}$Bo\u{g}aziçi
University}

\date{07.10.2016}
\begin{abstract}
We study 100 images of early LHC collisions that were recorded by
the ATLAS experiment and made public for outreach purposes, and extract
the charged particle multiplicity as a function of momentum for proton-proton
collisions at $\sqrt{\ensuremath{s}}=7$$\,$TeV. As the collisions
we study have already been pre-processed by the ATLAS Collaboration,
the tracks are visible, but are available to the public only in the
form of low-resolution bitmaps. We employ two separate image processing
methods, one based on the industry-standard OpenCV library and C++,
another based on self-developed algorithms in Python. We present the
transverse momentum and azimuthal angle distributions of the particles
obtained through both methods, in agreement with the literature.
\end{abstract}
\maketitle

\section{Introduction}

In 2009, when the Large Hadron Collider (LHC) at CERN started to deliver
its first physics-quality proton-proton collisions at a center-of-mass
(CoM) energy of $\sqrt{s}=900$$~$GeV, the very first measurements
by the ALICE, ATLAS and CMS Collaborations were of the charged particle
multiplicity$~$\cite{aliceFirst,atlas2009,cms2009}. The goal of
such measurements is to understand the momentum and angle distributions
of charged particles that result from the so-called \emph{minimum
bias events}, which are mostly inelastic collisions at low momentum
transfer between the constituents of the protons. Minimum bias events
have been useful in understanding the performance of the detectors,
and their detailed Monte Carlo simulations play a role in modelling
the background to signals of searched particles, such as the Higgs
Boson.

The aim of the present work is to perform this multiplicity measurement
with very limited resources and using data made public by the ATLAS
Collaboration in the form of digitised pictures of select collisions.
Since as early as 2010, the ATLAS Collaboration has been publishing
event displays of a fraction of the LHC collisions live from a publicly
available website$~$\cite{atlasPublicDisplay}. We captured 100 such
event displays in May 2011, during the so-called Run1 of the LHC,
with collisions at a CoM energy of $\sqrt{s}=7$$\,$TeV. The trajectories
of the charged particles have been reconstructed from the hits they
leave in the ATLAS inner detector and are seen as the tracks in the
event displays.

We describe two alternative methods for processing the collision images
to identify the tracks and extract their azimuthal angles and transverse
momenta. One method attempts to use an external image processing library,
while for the other, we develop our own application-specific algorithms.
In both cases, only free and open-sourced tools and software are used.
The programs we develop are portable and work on multiple operating
systems, as the authors had access to different resources. The two
methods are independent and set different templates for future educational
research projects of similar scope. All the development work has been
performed by young researchers with no or very limited programming
experience.

\section{Event D\i splays}

\begin{figure}
\includegraphics[width=0.85\linewidth]{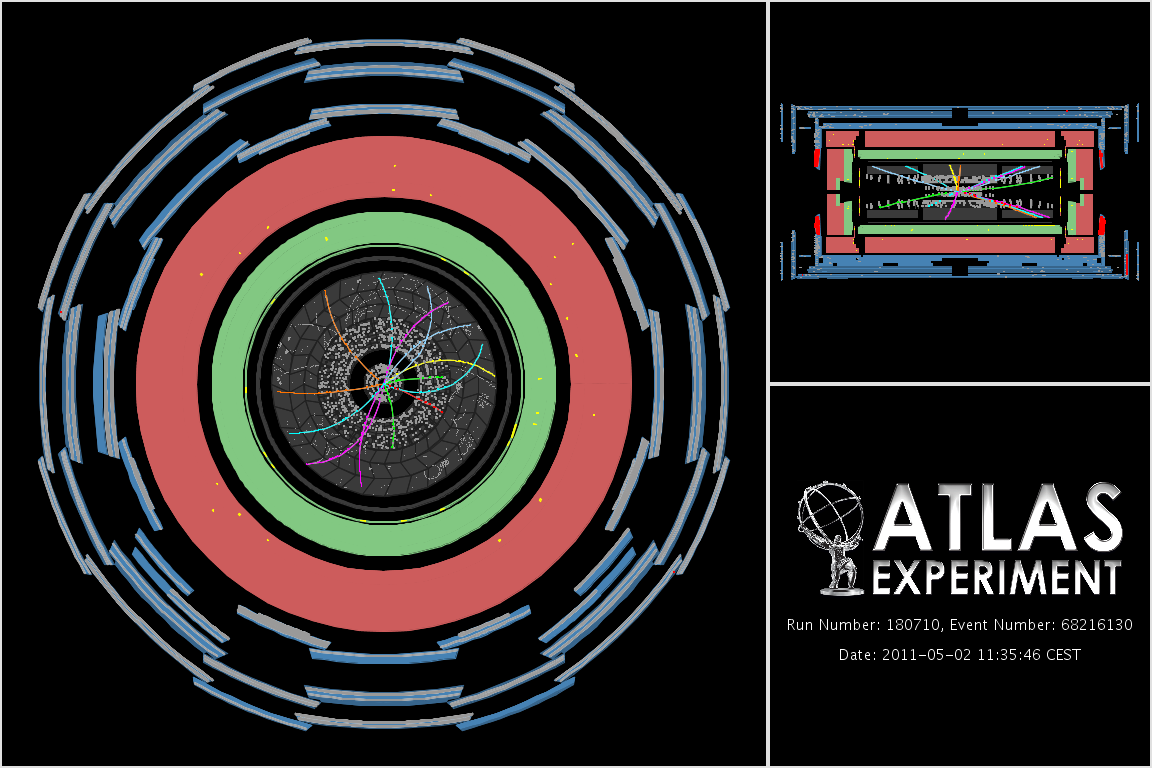}

\caption{An example event display from our data sample. In the absence of any
information to the contrary, we assume that the +$x$ (+$y$) axis
is pointing rightward (upward) in the transverse view. This is a low-occupancy
event with only 14 tracks.\label{fig:eventDisplay} }
\end{figure}

Our data is in the format of png files with $1152\times768$ resolution.
As seen in the example in Figure$~$\ref{fig:eventDisplay}, each
display has three panels. A large square panel ($765\times765$) on
the left side shows a projection of the detector transverse to the
beam ($z$) axis. ATLAS uses a right-handed coordinate system, with
the $+y$ axis pointing vertically upwards and the $+x$ axis pointing
towards the center of LHC ring$~$\cite{atlasDetector}. The top one
of the two smaller panels on the right is a $\rho$-$z$ projection
of the detector ($\rho$ is the radial distance to the $z$ axis),
while the bottom panel provides details about the particular event,
such as its date and time of collection.

The particles that originate from around the interaction point of
the beams at the center of detector spread outward going through the
inner detector (ID) first, then enter the calorimeters (green and
red concentric shells in the transverse view, with yellow boxes representing
energy deposits), and finally those that are not absorbed in the calorimeters
go through the muon chambers (blue thin blocks in the displays, with
gray bars or red points representing particle hits). The outlines
of the ID, which consists of three layers of silicon pixel detectors,
four more layers of silicon strip detectors, and about 30 layers of
gas straw detectors, are depicted as dark gray regions in the center
part of the transverse-view displays. These dark gray outlines are
clearly visible only for relatively ``empty'' events, as for most
events, the noise and particle hits in light gray color occupy most
of the ID in the transverse view.

The ID is immersed in a 2T magnetic field parallel to the $z$ axis.
This magnetic field bends the trajectories of the particles, dependent
on the magnitude of their transverse momenta, $P_{T}=\sqrt{{P_{x}^{2}+P_{y}^{2}}}$.
The resulting helical tracks, as reconstructed by the ATLAS software
from the ID hits, can be seen as colored circular arcs in the transverse
view, and as straight lines in the $\rho$-$z$ projection. Based
on displays with very few tracks, it is possible to deduce that each
track is drawn in the same color in the two views, but due to a limited
selection of track colors (7 in total), most often it is not possible
to unambiguously match the arcs to the straight segments. Because
of this, we choose to focus only on the higher-resolution transverse
view; forsaking the possibility of measuring the pseudorapidity, $\eta$,
of the particles for now. (Pseudorapidity is defined as a function
of the angle, $\theta$, of the particle with respect to the $+z$
axis: $\eta=-\ln(\tan(\theta/2)$.)

As the event displays do not come with any absolute length scale,
it needs to be extracted manually. It is worth noting that various
layers of the detector do not appear to be up to the same scale, for
instance the outermost muon chambers are much larger than what the
displays would imply. In order to make sure that at least the various
parts of the ID are on the same scale, we measure the apparent sizes
of various structures (like the diameters of the various subdetectors
of the ID) in pixel units and compare them with their actual sizes
from the literature. All these measurements yield ratios within a
couple percent of each other. As the best measurement, we pick that
of the outer diameter of the thin superconducting magnet, which appears
as a thin dark gray ring between the ID and the calorimeters. The
scale we obtain is exactly 256 pixels to 2.56$\,$m, so each pixel
is taken to represent 1$\,$cm.

\section{Identification of Track Colors}

\begin{figure}
\includegraphics[width=0.28\paperwidth]{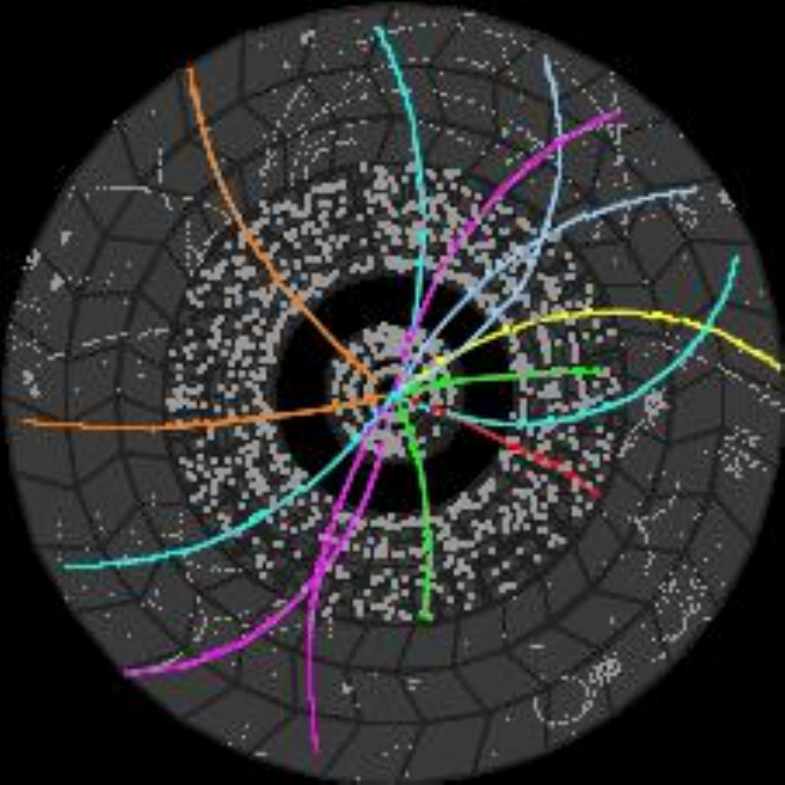}\includegraphics[bb=-50bp 50bp 621bp 650bp,width=0.33\paperwidth]{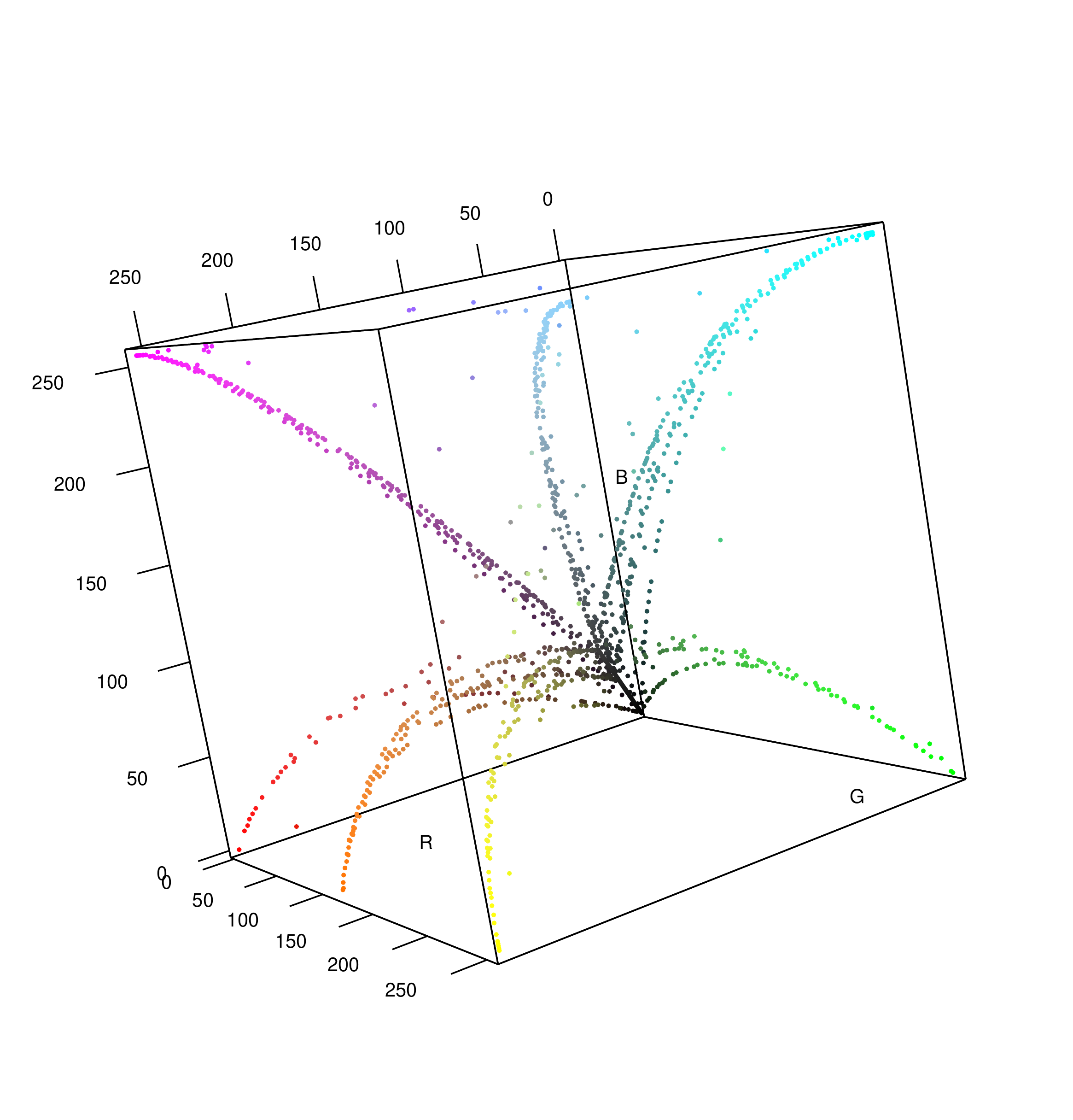}

\includegraphics[bb=20bp 20bp 921bp 769bp,width=0.4\paperwidth]{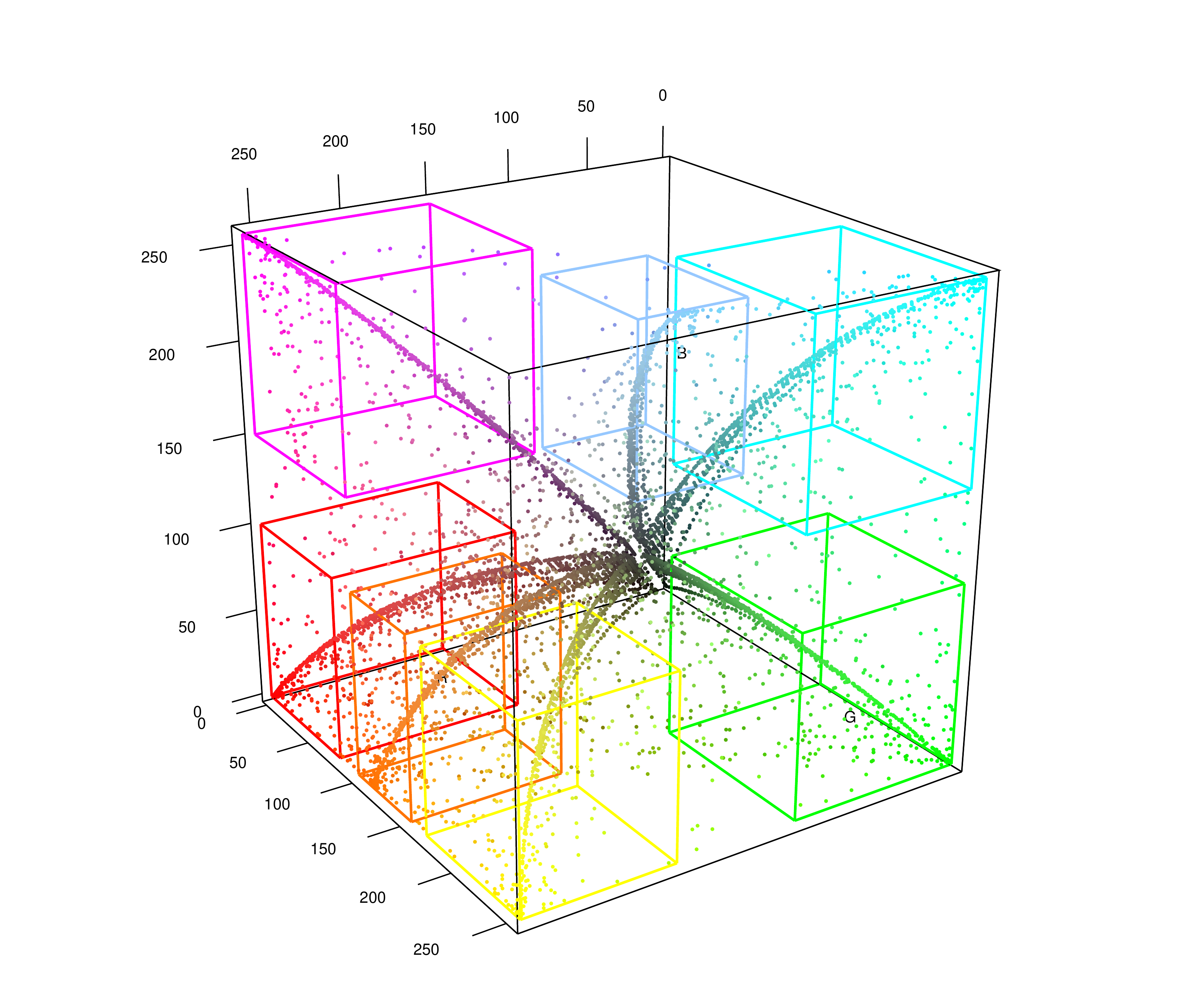}\includegraphics[bb=50bp 0bp 567bp 385bp,width=0.5\linewidth]{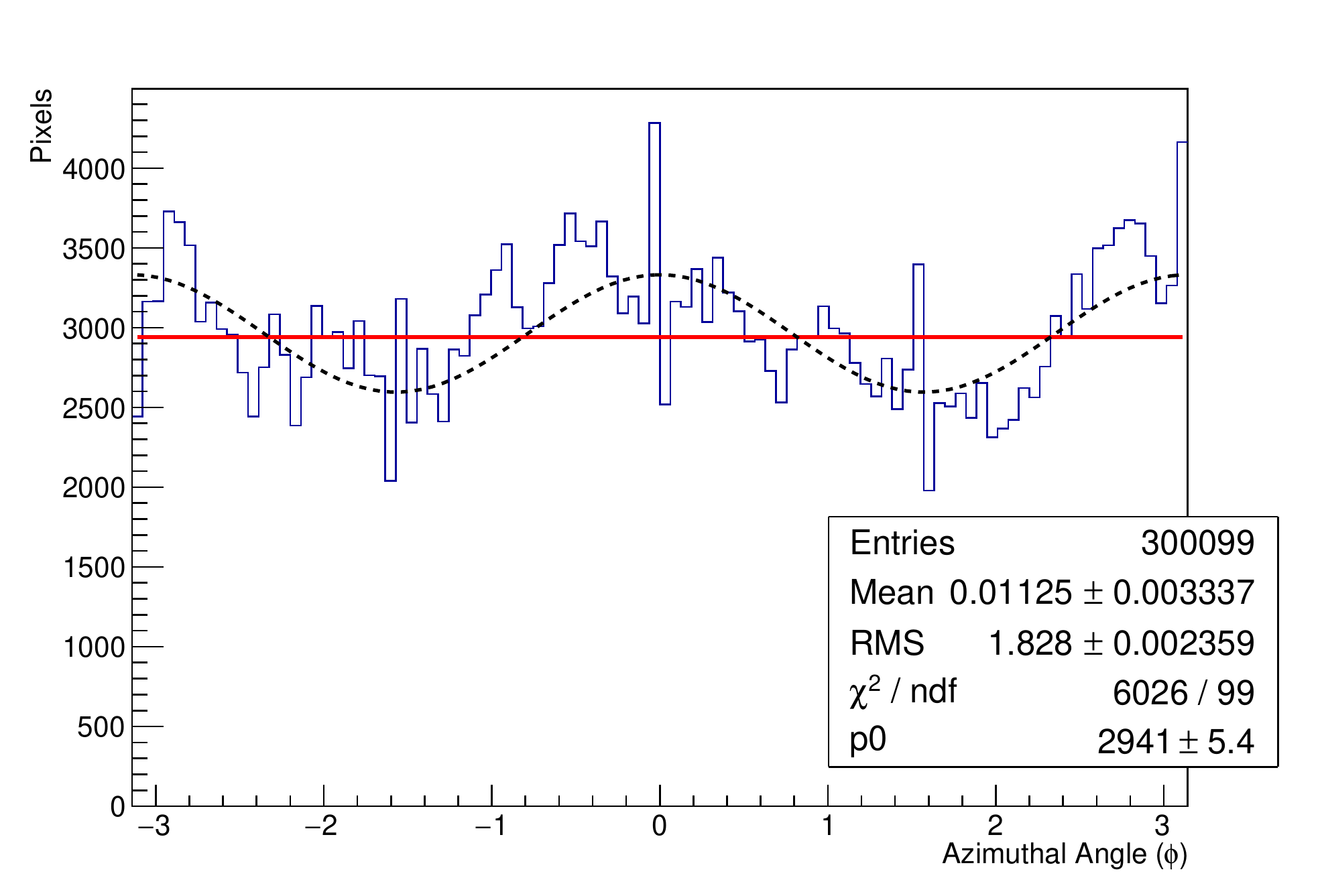}

\caption{Top left: The ID part of the example event display from Figure$~$\ref{fig:eventDisplay}.
Top right: The location in the (R, G, B) colorspace of all the pixels
from the top left image. The dark points around the origin (far bottom
corner of the cube) are due to the gray color of the ID outline. Seven
distinct ``threads'' can be perceived for each of the colors used
for the particle tracks. Bottom left: Colorspace shown for pixels
obtained from four high-occupancy events. The gray pixels have been
vetoed as described in the text. The colored rectangular prisms demonstrate
the selection cuts used to classify the pixels into one of the seven
colors. Bottom right: Azimuthal angle distribution of the pixels assigned
to any one of the colors. A flat line (red) has been fit to the histogram
to check the uniformity of the distribution. Also shown is a fit (black
dashed) to a function of the form ${\cal {N}}\times(1+\alpha\cos(2\phi))$,
which describes the distribution better.\label{fig:rgbSpace}}
\end{figure}

Both of our analyses start with a common part: the identication of
the distinct colors used for the tracks. For this purpose, we initially
remove all the pixels (by painting them black) in the images except
those that fall within the transverse view of the ID. Next we identify
where in the (R, G, B) colorspace, the seven track colors (red, dark
orange, yellow, lime, light sky blue, aqua and fuchsia$~$\cite{colorsCSS})
and the detector-related gray colors occupy. Towards this goal, we
scan the ID part of a few low-occupancy images, and look for clusters
in the (R, G, B) space, as seen in the top panels of Figure$~$\ref{fig:rgbSpace}.
The gray pixels lie close to the diagonal line between the origin
and (255, 255, 255), so in order to remove them, we veto any pixels
whose R, G, B coordinates are all within 4 points of each other.

As the seven track colors occupy distinct regions in the (R, G, B)
space, we are able to define identification criteria for each of them
with relatively high efficiency. We use simple 3D rectangular box
cuts (as in the OpenCV's \texttt{inRange()} method), with the boundaries
chosen in such a way that the selection efficiency of all the colors
are within about 10\% of each other. The loosest cuts (hence the highest
efficiency) are for fuchsia and lime, as they are at two corners away
from the other colors, while the tightest cut is for yellow, as its
darker hues get gradually similar to orange. The overall efficiency
of assigning the non-black, non-gray pixels to any one of the 7 colors
is about 69\% as measured in the ID regions of all 100 images in our
dataset.%
{} A significant fraction of the pixels lost during color identification
is mainly because of colors getting mixed up around the intersecting
tracks. While we note that this fraction will be dependent on the
occupancy of the ID, we do not attempt to correct for this systematic
effect, as our data consist of relatively-low-occupancy events from
the early LHC collisions, when the average collisions per each crossing
of the proton bunches was about 7$~$\cite{atlasLumi}.

Having identified the colors, we look for possible biases in our dataset.
Given the cylindrical symmetry of the ATLAS Detector, the tracks are
expected to be uniformly distributed over the whole range of azimuthal
angles (polar angle in the $x$-$y$ plane). The bottom right panel
of Figure$~$\ref{fig:rgbSpace} shows the histogram of the azimuthal
angle for all the pixels that we identify with any of the 7 colors.
We observe a small, but statistically significant bias: the distribution
is not really flat, as demonstrated by the very high value of the
$\chi^{2}$ of a straight line fit.%
{} We could not identify the cause of this bias; its magnitude and features
are partially dependent on color. However it is small enough to be
treated as a systematic uncertainty and we derive a correction weight
for it at the end of our analyses.%

\section{Analys\i s Us\i ng a Computer V\i s\i on L\i brary}

The first method we utilize to extract information from our data makes
use of a freely available computer vision library, OpenCV 2.4.6, which
supports hundreds of algorithms for pattern recognition$~$\cite{opencv}.
We chose OpenCV mainly because of its sizeable user community, providing
a starting programmer with the avenue of quick online feedback and
a large number of code examples. None of the authors had any prior
experience with OpenCV or other computer vision libraries. One of
the authors, GÜ, tried the C bindings of the library, but we quickly
decided that the C++ bindings were more natural and easy to work with.
It is worth noting that at this stage, GÜ had no formal training with
either C or C++, or any other programming language, beyond some limited
tutoring by VEÖ. Our development environment was Orwell Dev-C++ on
Windows 8$~$\cite{devcpp}.

\begin{figure}
\includegraphics[width=0.32\columnwidth]{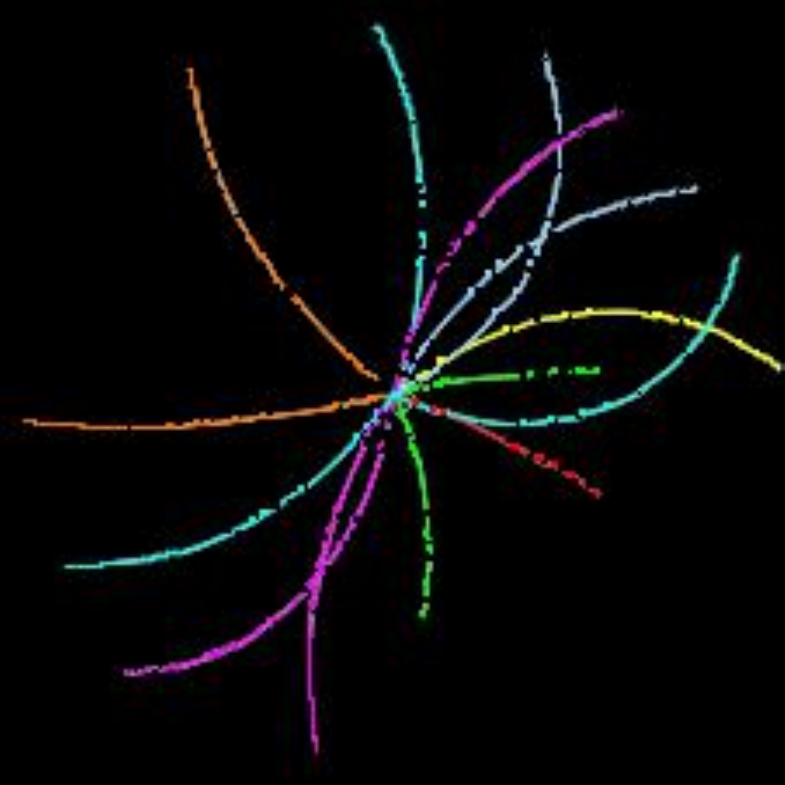} \includegraphics[width=0.32\columnwidth]{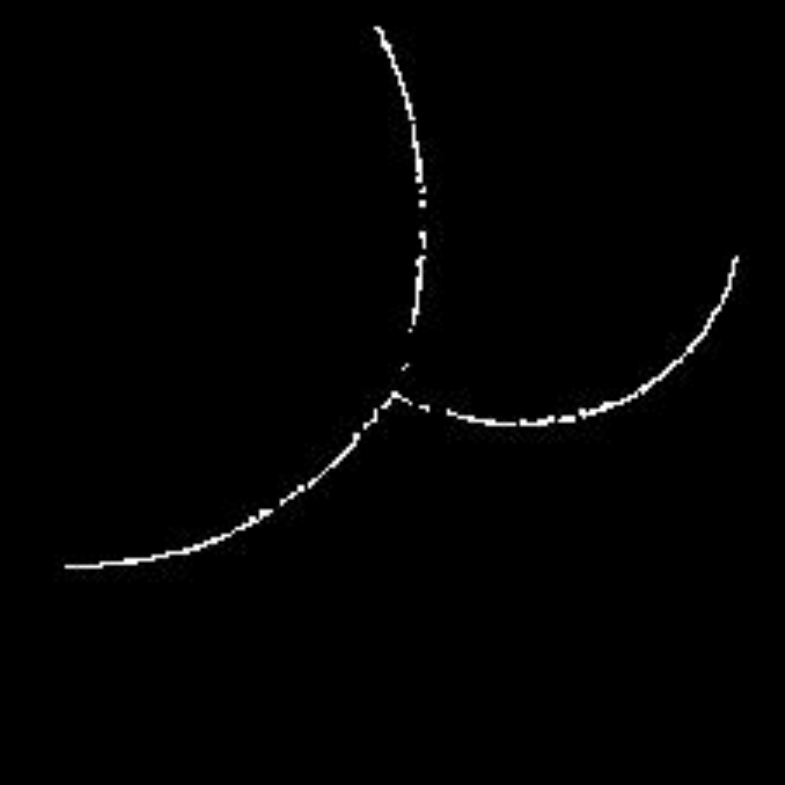}
\includegraphics[width=0.32\columnwidth]{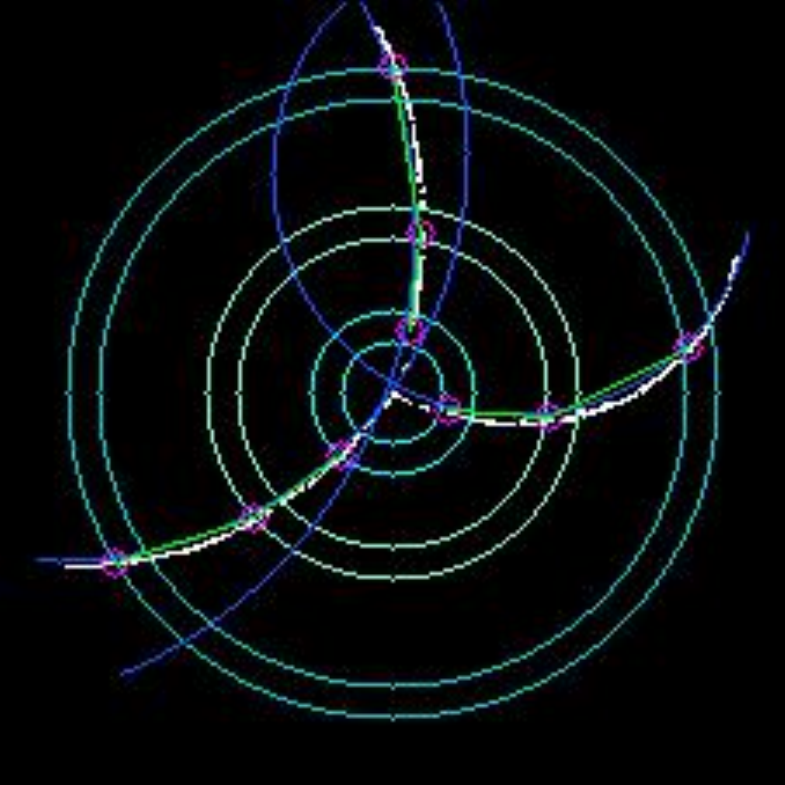}

\caption{Steps of the ``combinatorial'' analysis, applied to the example
event from Figure$~$\ref{fig:eventDisplay}. Left: Background pixels
cleaned. Middle: Pixels of a given color (for example, aqua) selected.
Right: Three rings defined; in each ring pixels selected; and from
triplets of pixels, tracks reconstructed.\label{fig:steps-of-opencv-analysis}}
\end{figure}

For the computer-assisted identification of the tracks, various out-of-the-box
techniques from OpenCV were tried. Some of these included connected-component
analysis, various generic feature extraction functions such as \texttt{goodFeaturesToTrack()},
and the Hough transform (with various degrees of Gaussian smoothing)
for extracting circular arcs$~$\cite{goodFeaturesToTrack,houghCircle}.
Unfortunately these attempts yielded a high rate of mis-reconstructed
circles and low efficiency. Furthermore, the performance was highly
dependent on the detector occupancy. So a decision was taken to implement
an algorithm of our own.

As a preparatory step, we wrote a small program with a graphical user
interface that allows an end-user to open any image and pick pixels
by clicking with a mouse. With each pixel selected, the program prints
out its spatial as well as its colorspace coordinates. When a set
of three pixels (triplet) is selected, the program solves for the
equation of the circle that passes through them, draws the circle
and its center on the screen, and prints the calculated $P_{T}$.

Our algorithm attempts to automate the manual procedure of picking
triplets of same colored pixels. We first split the background-cleaned
image into seven subimages, by picking the pixels of each of the identified
colors one at a time. On these subimages, we define three 9-pixel-thick
concentric rings (of inner radii 15, 45 and 85 pixels, covering parts
of the silicon pixel, microstrip and gas straw detectors respectively)
and make a list of the colored pixels that fall into them (Figure$~$\ref{fig:steps-of-opencv-analysis}).
Whenever two pixels in our list are very close, we choose one randomly.
Here we measure the closeness of two pixels as follows: Let $\vec{A}$
and $\vec{B}$ represent the 2D position vectors of the pixels with
respect to the center of the ID. Our measure of closeness, which we
term ``azimuthal distance'', is given by $|\vec{A}\times\vec{B}|/|\vec{A}|$,
which is the length of the component of $\vec{B}-\vec{A}$ that is
perpendicular to $\vec{A}$. For pixels in the inner, middle, and
outer ring, we require that the azimuthal distance between any two
pixels to be at least 6, 8, and 10 pixel sizes, respectively.

With our list of pixels, we try combinations of triplets of pixels,
one from each ring, and apply constraints on the locations of the
three hits with respect to each other. These constraints are chosen
as loosely as possible in order to have a unifom reconstruction efficiency
as a function of track $P_{T}$. Each pixel from the middle or outer
rings is allowed to appear in one triplet only, while pixels from
the inner ring can appear in more than one triplet, in order to account
for the high occupancy of hits close to the center of the ID. While
this decision occasionally generates some partially mis-reconstructed
tracks, we find the rate of fakes acceptable in return for higher
reconstruction efficiency. Finally, we solve for the equation of the
circle for our surviving triplets.

\section{Analysis Using Application-Specific Algorithms}

The goal of our alternative analysis was to perform a detailed pixel-by-pixel
processing with completely home-grown algorithms. This would not only
serve as a fully-independent cross-check to our OpenCV analysis, but
also as a vehicle for testing new algorithm ideas that are specifically
tuned for the available data. The work was performed initially with
Python$~$2.6 on Windows, and later with Python$~$3.4 on Linux and
OSX, by two of the authors, EBa and EBe, who had some previous programming
experience (but neither had any experience with the advanced features
of Python such as classes, iterators, decorators, functional programming,
etc.). The code is standard Python with two external dependencies:
SciPy 0.13, which is used for least squares optimization, and Pillow$~$3.3.1,
the fork of the lightweight Python Imaging Library (PIL), which is
solely used to read the input files and manipulate them pixel by pixel$~$\cite{scipy,pil,pillow}.

\begin{figure}
\includegraphics[width=0.32\columnwidth]{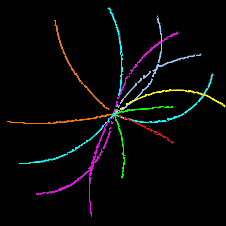} \includegraphics[width=0.32\columnwidth]{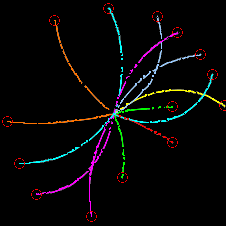}
\includegraphics[width=0.32\columnwidth]{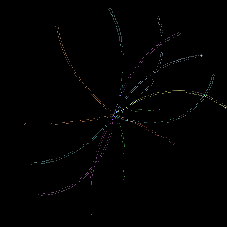}

\caption{Some of the steps of the ``clustering'' and ``sweeping'' analyses
being applied to the example event from Figure$~$\ref{fig:eventDisplay}.
Left: Outcome of tracklet-merging and filling of the missing parts.
Middle: Outer endpoints of the outmost tracklets identified. Right:
The unused pixels that were discarded or missed by the sweeping algorithm.
They appear as weak shadows of the reconstructed tracks.\label{fig:stepsOfPILanalyses}}
\end{figure}

The first algorithm we devise is a clustering algorithm. We start
by scanning the ID in search for the pixels that are at the outer
endpoints of the circular arcs. Hereafter we use the prefix in- (out-)
to describe closer-to (further-away-from) the center of the ID. Hence
the outer endpoints of the arcs are pixels of any one of our 7 colors,
with no other neighboring pixels of the same color that is further
away from the center of the ID. For each outer endpoint found, we
start our outside-in clustering: a tracklet is formed as a set of
pixels of a given color, which are inward neighbors of the endpoint
or the other pixels in the set. The next step is the merging of tracklets
into tracks (Figure$~$\ref{fig:stepsOfPILanalyses}, left panel).
The intersection of different colored tracks or the presence of gray-colored
detector hits break the tracks into pieces, which get reconstructed
as separate tracklets. We identify the innermost pixel of each tracklet
and check if the outer endpoint of another one of the same color is
close by. As a final step, we handle incorrectly clustered parts of
intersecting tracks of a given color, by searching for kinks in our
clusters by examining unexpected jumps in the azimuthal angles of
the pixels. For such cases we rerun our clustering algorithm with
constraints on the angles. The result is satisfactory for the example
event from Figure$~$\ref{fig:eventDisplay}, where we are able to
reconstruct all 14 tracks.

Unfortunately, the performance degrades quickly with increased multiplicity
of tracks. The tracklet merging and the kink identification have a
number of parameters that need careful tuning and we have observed
that high occupancy events require highly different tunes. As an alternative
solution, we devise a second algorithm, the ``sweeping'' analysis.
From the clustering algorithm, we take the best performing part, outer-endpoint
identification, and perform the following steps:
\begin{enumerate}
\item Apply endpoint identification to find the outer endpoints of the outmost
tracklets (Figure$~$\ref{fig:stepsOfPILanalyses}, middle panel).
\item For each identified end-point, define a 17-pixel-thick ring that is
centered on the endpoint and whose mid-radius is half the distance
between the endpoint and the center of the ID. Make a set of ``candidate
midpoints'': pixels that fall into the ring and whose color is the
same as the endpoint.
\item For each midpoint identified in the second step, define a circular
arc that connects the endpoint to the center of the ID and passes
through the midpoint. Count how many other pixels of the same color
are on or within $\pm2$ pixels around each such arc.
\item Amongst the arcs defined in the third step, pick the one with the
highest count of pixels. Perform a least squares fit (details below)
to those pixels to extract the precise equation of the arc. From its
radius, compute the transverse momentum $P_{T}[\mathrm{GeV/}c]=B[\mathrm{T}]\times r[\mathrm{m}]\times0.2998[\mathrm{m/ns}]$
and other parameters of interest. Remove the used pixels from the
original image.
\item Repeat steps 2 thru 4 until no tracks are left to be reconstructed,
``sweeping out'' a very large fraction of all the colored pixels.
The few remaining pixels at the end of the whole procedure can barely
be perceived in the rightmost panel of Figure$~$\ref{fig:stepsOfPILanalyses}.
\item On rare occasions, a track gets reconstructed more than once using
mutually exclusive subsets of its pixels. To remove the duplicates,
search for pairs of reconstructed arcs of the same charge (based on
whether they turn clockwise or counterclockwise), of the same color,
and whose centers are within 15 pixels of each other. If the associated
number of pixels for either of the pair is fewer than 15, or if the
measured radii of the pair are within 0.5\% of each other, discard
the arc with the fewer pixels.
\end{enumerate}
The circle fit we perform in step 4 itself proceeds through two substeps.
The first substep has been implemented by one of the authors, EBa,
who derived an algebraic solution using basic principles from a linear
algebra textbook$~$\cite{gilbertStrang}. This solution looks for
values of $(x_{C},y_{C},R)$, coordinates of the center and the length
of the radius, that minimize $S=\sum_{i}\left((x_{i}-x_{C})^{2}+(y_{i}-y_{C})^{2}-R^{2}\right)^{2}$,
where $(x_{i},y_{i})$ are the coordinates of the pixels being fit.
A later literature search revealed that this minimization is known
as Kåsa's method$~$\cite{kasa}. Unfortunately, a visual inspection
quickly reveals that this method systematically underestimates the
radius for incomplete circles like the arcs we have in our data. So
as a second substep, we proceed to do a full least squares fit that
minimizes $S^{\prime}=\sum_{i}\left(\sqrt{(x_{i}-x_{C})^{2}+(y_{i}-y_{C})^{2}}-R\right)^{2}$
using numerical methods. For this purpose, \texttt{optimize.leastsq()}
function from the SciPy library, an implementation of the Levenberg-Marquardt
algorithm, is used. We pass the center coordinates computed using
Kåsa's method as an initial estimate to the numerical method, in an
attempt to improve the speed of the computation.

\begin{figure}
\includegraphics[width=0.47\columnwidth]{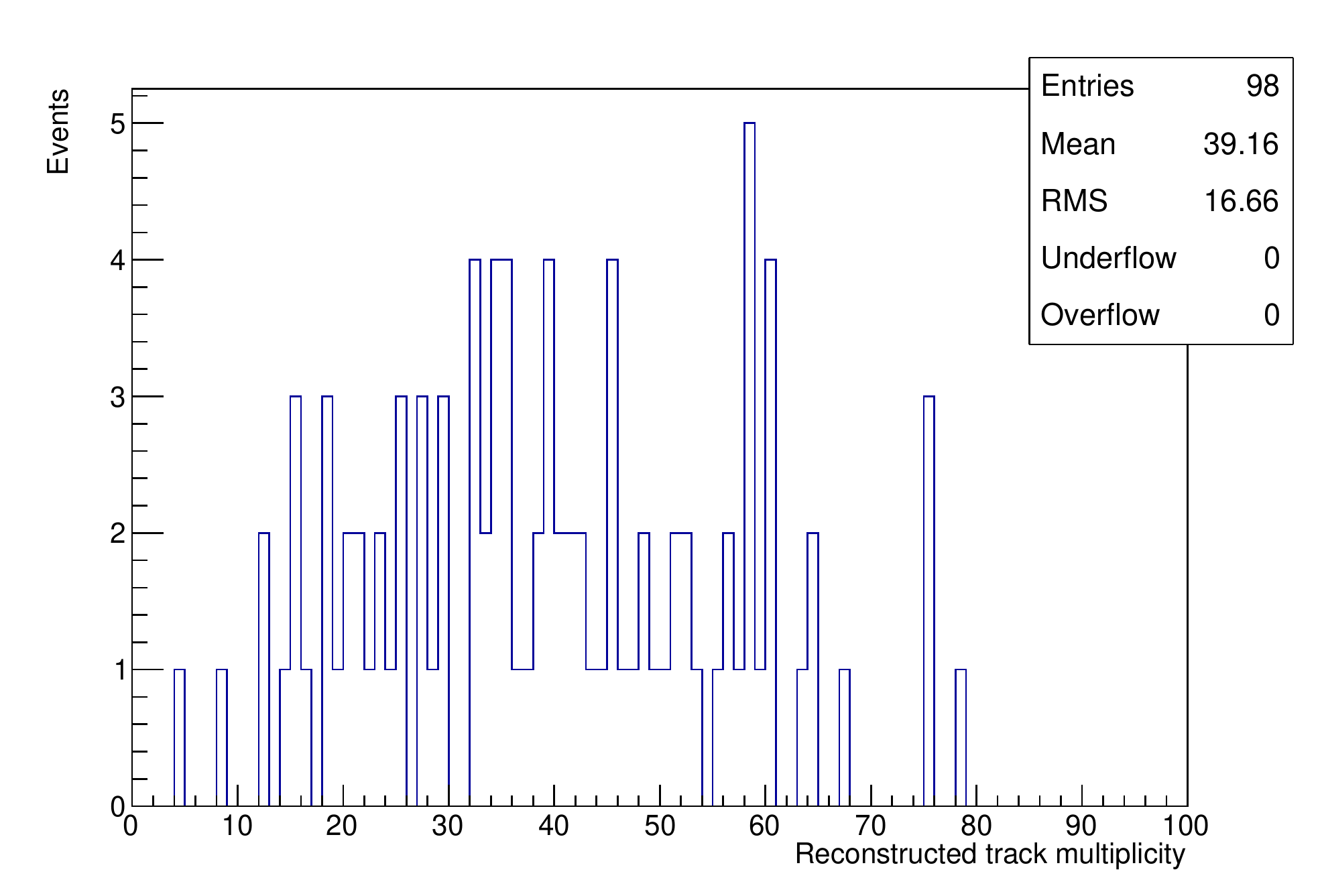}\includegraphics[width=0.47\columnwidth]{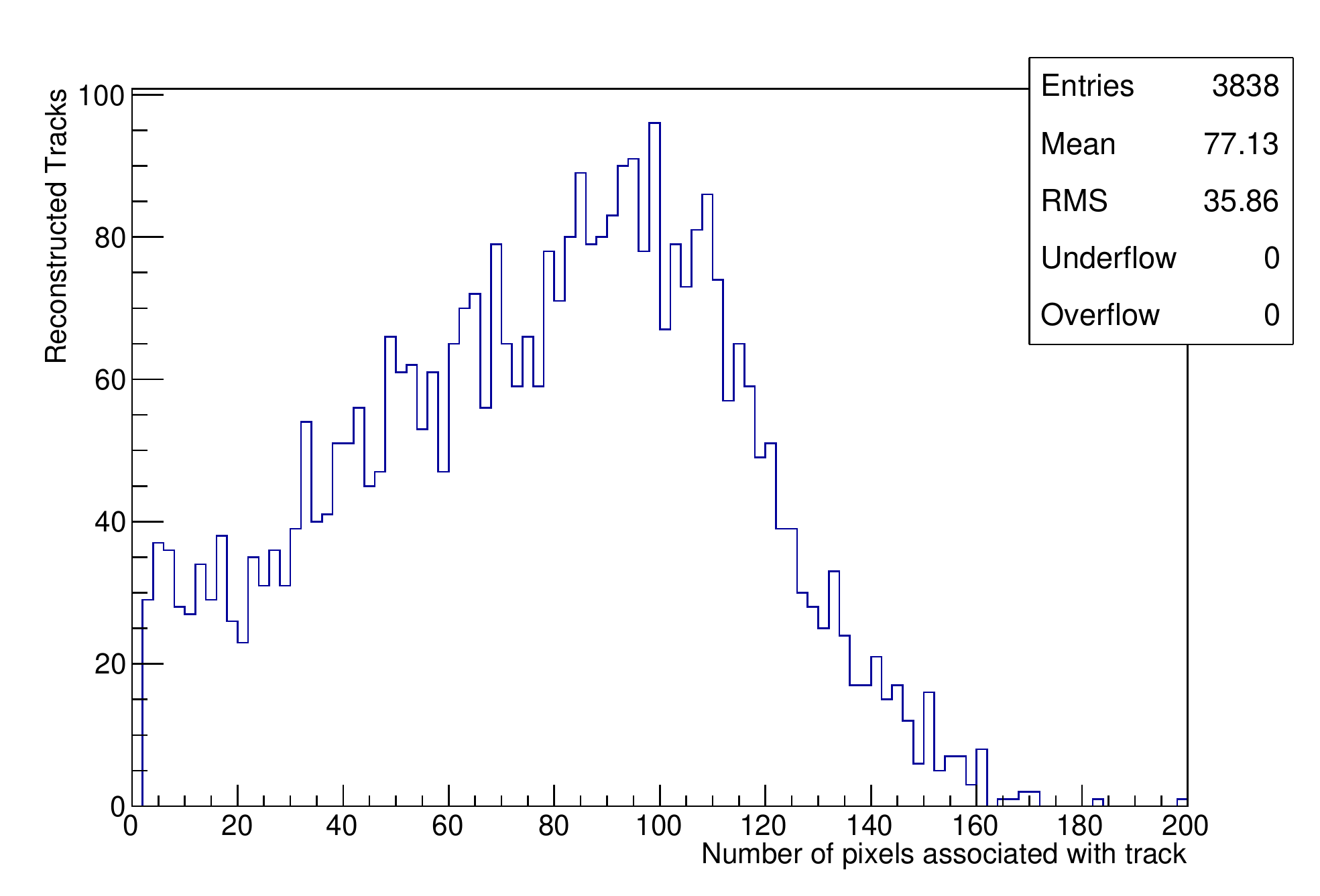}

\includegraphics[width=0.47\columnwidth]{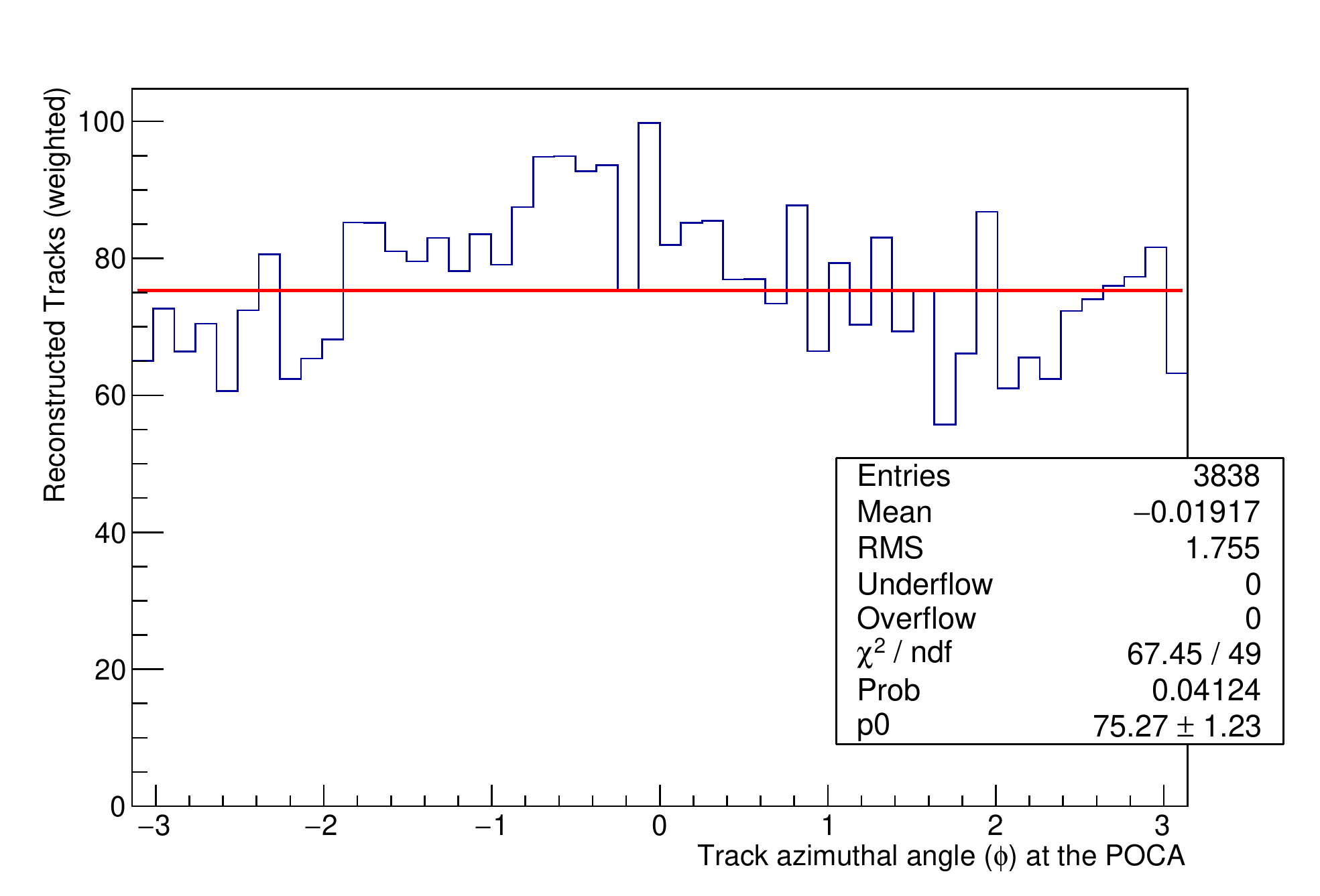}\includegraphics[width=0.47\columnwidth]{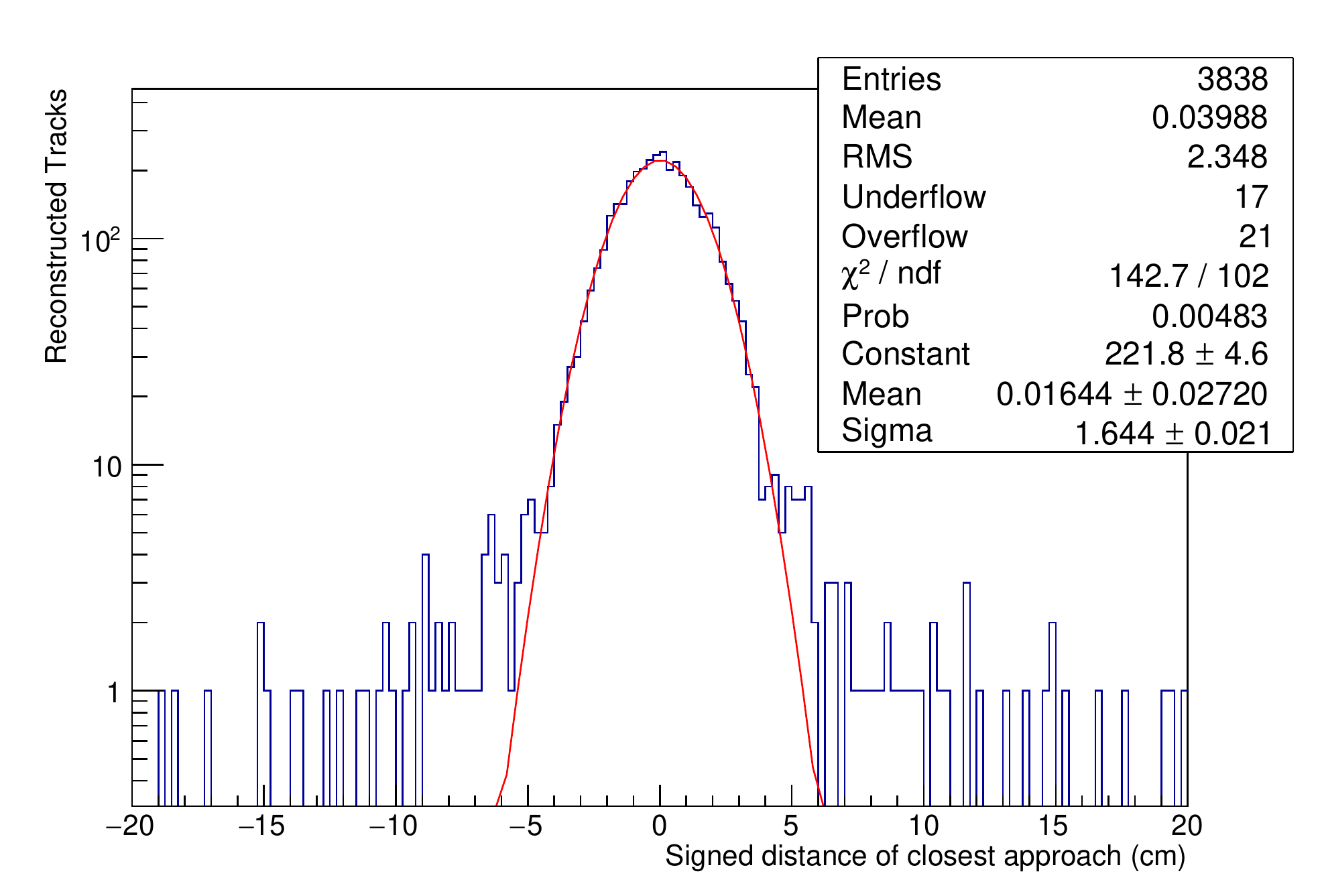}

\caption{Some results from the ``sweeping'' analysis. Top left: Track multiplicity
per event. Top right: Number of pixels associated with the reconstructed
tracks. Bottom left: Azimuthal angle of tracks computed at the point
of closest approach to the center of the ID. The histogram is filled
with each track assigned a weight given by $(1+\alpha\cos(2\phi))^{-1}$,
where $\alpha=0.124$, in order to overcome the bias observed in the
distribution of the pixels in Figure$~$\ref{fig:rgbSpace}. A flat
line fit (in red) is also shown. Bottom right: Distance of closest
approach to the center of the ID, multiplied by the charge of the
track. A Gaussian function (in red) has been fit to the histogram.\label{fig:sweepingAnalysisResult} }
\end{figure}

The final results from the sweeping analysis are shown in Figure$~$\ref{fig:sweepingAnalysisResult}.
Over 3800 tracks have been reconstructed from 98 events with an average
number of 77 pixels per track. (Two events in our dataset had significant
activity only in the forward muon chambers and no ID tracks.) After
a visual inspection of the busiest images (those with over 60 tracks),
we conclude that our reconstruction efficiency is close to 100\%.
(In comparison, the fast combinatorial analysis reconstructs about
60\% of those tracks; some lost because they do not reach the gas
straw detectors.) The fake rate (tracks mis-reconstructed from unrelated
pixels) is harder to estimate accurately; as a substitute we modify
some of the critical parameters of our algorithms to see how robust
our final results are. The parameter to which the sweeping algorithm
is most sensitive appears to be the width of the arc in third step.
Doubling ($\pm4$ pixels) this parameter means lumping more hits together
and decreases the total number of tracks by about 8\%. We believe
that assuming a fake rate of about 10\% is then reasonable.

The angular distribution of the reconstructed tracks is mostly flat,
after taking into account the bias observed in the distribution of
the pixels in the dataset. The resolution in the measurement of the
transverse impact parameter, ie. the transverse distance of closest
approach to the center of the ID, is determined to be 1.64$\,$cm,
higher than our naive expectations. Upon further investigation, we
observe that this value is a decreasing function of the number of
pixels associated with the tracks, as we would expect, but the relationship
is not a simple $\propto N_{pixels}^{-1/2}$ function, suggesting
that the pixelation of the images plays an important role: perhaps
the actual center of the ID is not really aligned with the center
of any given pixel. Such a hypothesis could also provide a partial
explanation of the bias in the azimuthal angle distribution of the
colored pixels, but our attempts to confirm the hypothesis have failed.%

\section{Comparison to Past Measurements}

As we do not have the pseudorapidity values of the tracks, a direct
comparison to the published ATLAS results from Reference$~$\cite{atlas2010}
is not possible. However, in that publication it is stated that the
ATLAS measurement is compatible with the simulated collisions from
the Monte Carlo event generator program Pythia6 after a special tuning
of the program's parameters. We use Pythia$~$6.4.28, with the advised
AMBT1 tune to generate 1000 collision events of our own for comparison
with our results$~$\cite{pythia6}. Unfortunately such events are
not directly comparable either, as the ineffiencies of the ATLAS Detector
need to be taken into account.

\begin{figure}
\includegraphics[width=0.47\columnwidth]{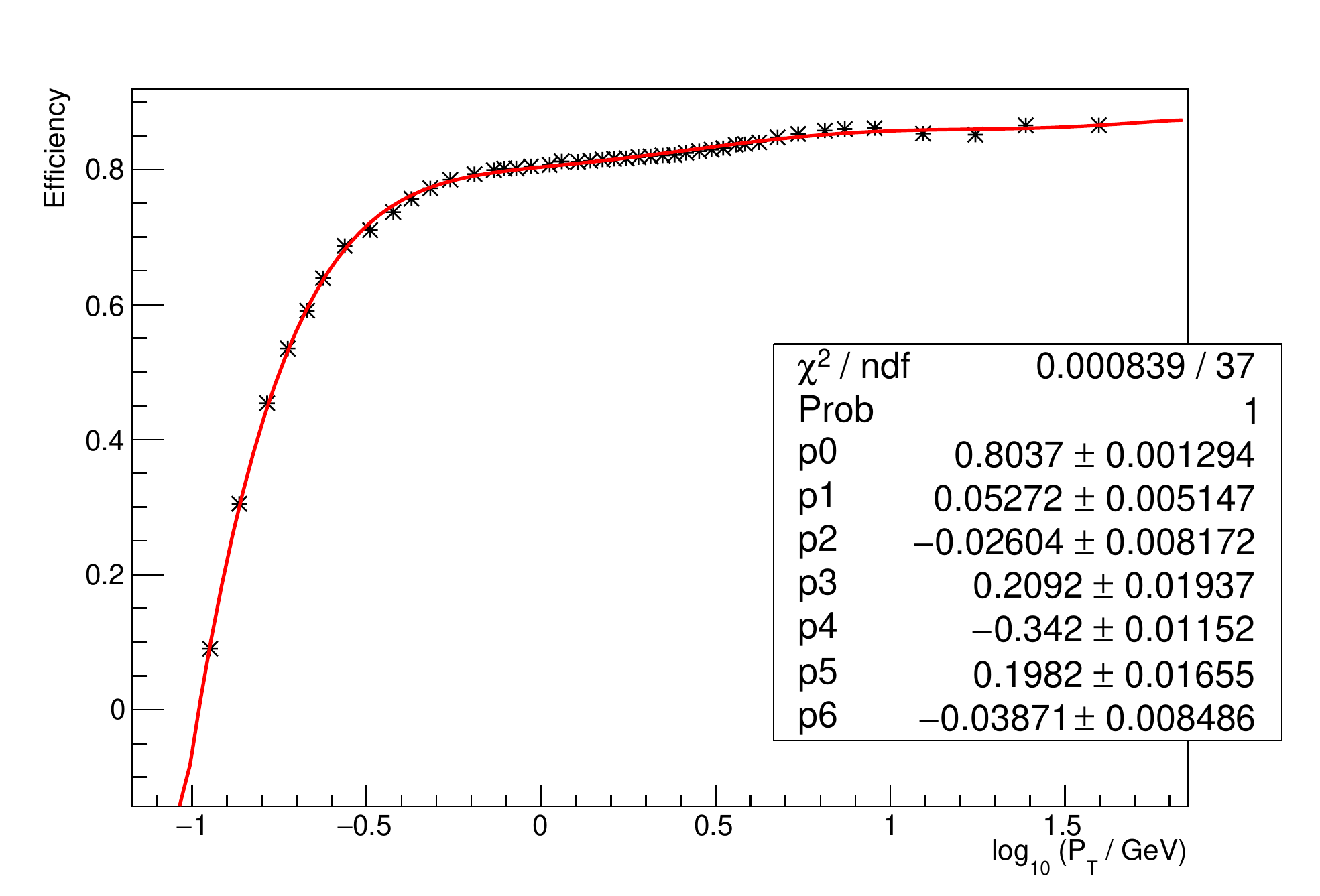}\includegraphics[width=0.47\columnwidth]{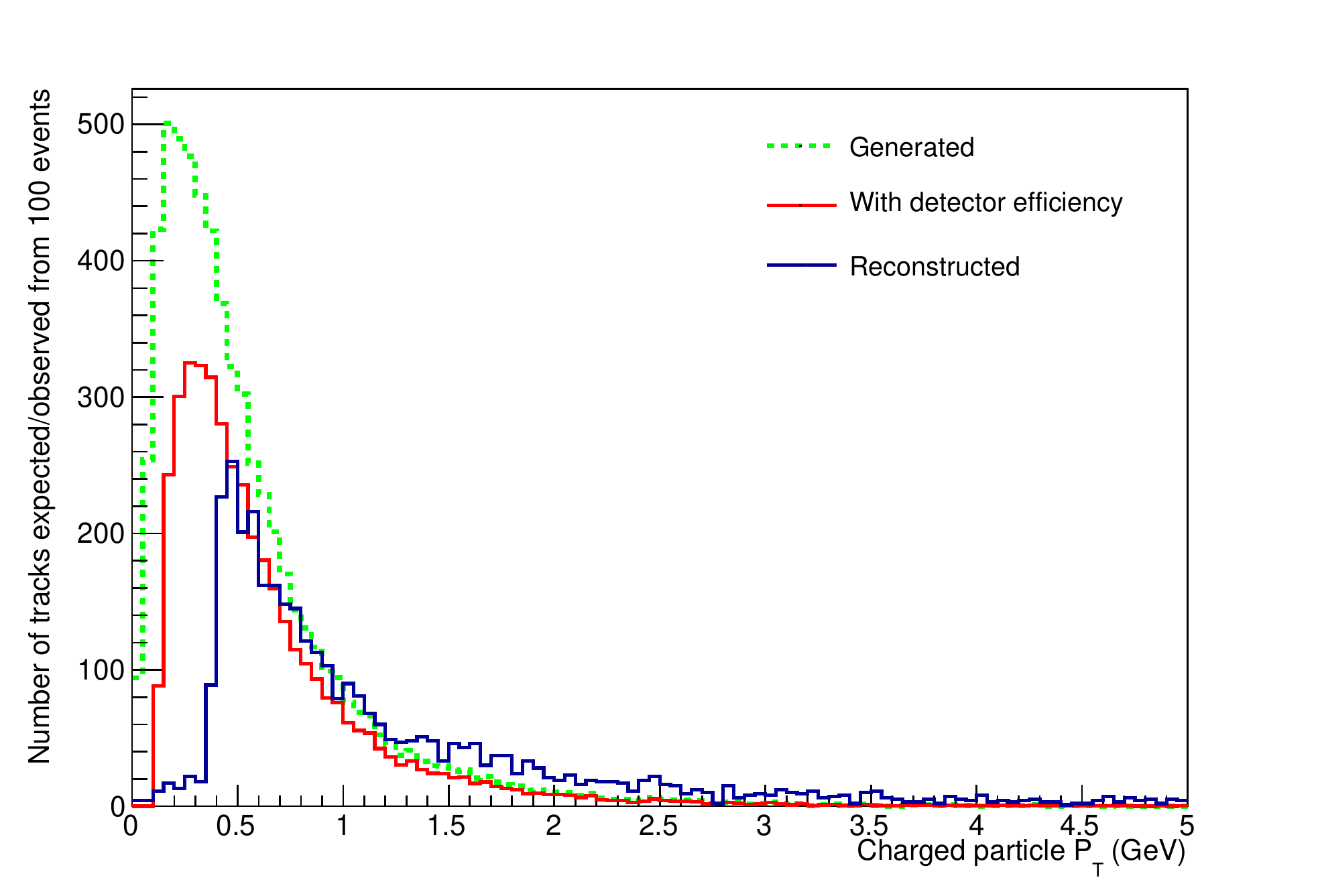}

\caption{Left: Efficiency of the ATLAS Detector as a function of $\log(P_{T})$.
Data points (black stars) have been extracted from Reference$~$\cite{atlas2010}.
The listed parameters $p_{0},...,p_{6}$ are the constants of the
fitted polynomial (red curve). Right: Comparison of the charged particle
transverse momentum histograms obtained as a result of our reconstruction
(blue solid) and from Monte Carlo simulation (red solid). Also shown
is the distribution from the MC generator before the $P_{T}$-dependent
ATLAS detector efficiency corrections have been applied (green dashed).\label{fig:comparisonToMC}}
\end{figure}

In order to imitate those inefficiencies, we digitize and extract
values from Figure 2(d) of$~$\cite{atlas2010} using the open-source
program WebPlotDigitizer$~$3.3$~$\cite{webplotdigitizer}. With
another open-source program, ROOT$~$5.34/24$~$\cite{root}, a 6th-order
polynomial is fit to the extracted values in order to obtain an empirical
formula describing the detector efficiency as a function of $\log(P_{T})$
(Figure$~$\ref{fig:comparisonToMC}, left). In the very low $P_{T}$
region where our empirical function gives negative efficiencies, the
efficiency is taken to be zero, while for $P_{T}>40$$\,$GeV/$c$
(above the range of the ATLAS plot), we take a constant value of 86.6\%.
For each charged particle in the Pythia6-generated events, the efficiency
corresponding to its $P_{T}$ can then be computed with our empirical
function and applied as a weight while filling the simulation histograms.

The comparison of the simulated charged particle $P_{T}$ and the
results of our sweeping analysis are shown on the right panel of Figure$~$\ref{fig:comparisonToMC}.
Higher momentum tracks are comparatively more frequent in our data,
but the most significant difference is the shortage of tracks below
0.5$\,$GeV/$c$. We suspect that both of these observations are related
to ATLAS's selection of events for public consumption; 0.5$\,$GeV/$c$
could be the side effect of the ATLAS trigger that feeds the data
stream used for outreach purposes. In our comparison plot, we have
assumed the presence of such a cutoff in determining the normalization
of the simulated data. Despite the minor differences mentioned, we
find the level of agreement unexpectedly satisfactory.

\section{Conclusion}

We have shown the feasibility of repeating a cutting-edge measurement,
one that was featured as the beginning of ``the dawn of exploration
of a new energy frontier at the LHC'' in a Physics Synopsis by the
American Physical Society$~$\cite{APSsynopsis}, with the help of
freely-available tools being used by young researchers with no or
little prior experience. (GÜ was a high school student for most of
her work on the OpenCV-based analysis, while EBe and EBa, who devised
the clustering and sweeping algorithms were junior and freshman physics
majors respectively.) Despite our lack of a full understanding of
the detector behaviour, our results are broadly in agreement with
the simulated results that have been generated in accordance with
the publications by the ATLAS Collaboration.

The two different approaches we have described are complementary and
constitute examples of projects of different scopes. The combinatorial
analysis is fast (processing 100 events in about 4 minutes on a 2.2
GHz Intel Core i7), utilizes compiled C++ code and an industry-standard
computer vision library, and does inside-out tracking. Its logic closely
resembles fast trigger tracking codes used by the LHC experiments.
The clustering and sweeping analyses are much slower (1 event per
minute on average), utilize interpreted Python code and a lightweight
image manipulation library, and perform outside-in tracking. It is
in the spirit of slow offline tracking algorithms of the LHC experiments,
and yields results of high resolution. In spite of the fact that Python
is arguably a more expressive language than C++, their code is longer
and more complex than that of the combinatorial analysis.

The displays of events we collected in the early days of the 7$\,$TeV
running of the LHC has given us a unique opportunity to study the
charged particle tracks in the inner detector of the ATLAS Detector.
The higher CoM energy and much higher instantaneous luminosity make
the possibility of repeating our work difficult with more recent collisions,
due to the large amount of particles in the events. However the new
fish-eye event displays made public by the ATLAS Experiment show energy
deposits in the calorimeters and hits in the muon chambers very distinctly,
allowing interested students to share the experience we have outlined
in this paper in their own unique ways. In such future studies, it
might be possible to measure missing transverse energy, a parameter
fundamental in the search for beyond the Standard Model physics.
\begin{acknowledgments}
We would like to thank Deniz Niyazi for her support of GÜ during the
middle stages of the project, and Fatih Mercan for providing feedback
on our early drafts. VEÖ would like to thank Bo\u{g}aziçi University
Foundation (BÜVAK) and the Science Academy Society of Turkey for their
indirect financial support. We are grateful for the access that the
LHC experiments provide to their data, in formats that are inspiring
for the public.\end{acknowledgments}


\begin{thebibliography}{10}
\bibitem{aliceFirst}ALICE Collaboration (K. Aamodt et al.), ``First
proton\textendash proton collisions at the LHC as observed with the
ALICE detector: measurement of the charged-particle pseudorapidity
density at $\sqrt{s}=900$$~$GeV,'' Eur. Phys. J. C \textbf{65},
111\textendash 125 (2010).

\bibitem{atlas2009}ATLAS Collaboration (G. Aad et al.), ``Charged-particle
multiplicities in pp interactions at $\sqrt{s}=900$$~$GeV measured
with the ATLAS detector at the LHC,'' Physics Letters B \textbf{688},
21\textendash 42 (2010).

\bibitem{cms2009}CMS Collaboration (V. Khachatryan et al.), ``Transverse-momentum
and pseudorapidity distributions of charged hadrons in pp collisions
at $\sqrt{s}=0.9$ and 2.36$~$TeV,'' JHEP \textbf{02} (041), 1-19
(2010).

\bibitem{atlasPublicDisplay}Live Collisions in the ATLAS Detector,
\url{http://atlas-live.cern.ch/public/}

\bibitem{atlasDetector}ATLAS Collaboration (G. Aad et al.), ``The
ATLAS experiment at the CERN Large Hadron Collider,'' JINST \textbf{3}
S08003, i-407 (2008).

\bibitem{colorsCSS}We use the names of the closest colors from the
CSS3 recommendation by W3C. T. Çelik, C. Lilley, L. D. Baron (ed.),
``CSS Color Module Level 3, W3C Recommendation 07 June 2011'', \url{https://www.w3.org/TR/css3-color/}

\bibitem{atlasLumi}ATLAS Collaboration, ``Luminosity Public Results,''
\url{https://twiki.cern.ch/twiki/bin/view/AtlasPublic/LuminosityPublicResults#2011_pp_Collisions}

\bibitem{opencv}Open Source Computer Vision (OpenCV), \url{http://opencv.org/}

\bibitem{devcpp}Orwell Dev-C++ provides a suite for running gcc on
the Windows platform. \url{http://orwelldevcpp.blogspot.com}

\bibitem{goodFeaturesToTrack}Jianbo Shi and Carlo Tomasi, ``Good
Features to Track,'' Proceedings of the IEEE Conference on Computer
Vision and Pattern Recognition, 593-600 (1994).

\bibitem{houghCircle}H.K.$~$Yuen, J.$~$Princen, J.$~$Illingworth,
J.$~$Kittler, ``Comparative study of Hough Transform methods for
circle finding,'' Image and Vision Computing \textbf{8} (1), 71-77
(1990).

\bibitem{scipy}Eric Jones, Travis Oliphant, Pearu Peterson and others,
``SciPy: Open source scientific tools for Python'', \url{https://www.scipy.org/}

\bibitem{pil}Python Imaging Library, PIL, \url{http://www.pythonware.com/products/pil/}

\bibitem{pillow}Pillow, The Friendly PIL Fork, \url{https://python-pillow.org/}

\bibitem{gilbertStrang}Gilbert Strang, \emph{Linear algebra and its
applications}, 2nd edition (Academic Press, New York, 1980), pp. 111-120.

\bibitem{kasa}I. Kåsa, ``A circle fitting procedure and its error
analysis'', IEEE Transactions on Instrumentation and Measurement
\textbf{25}, 8-14 (1976).

\bibitem{atlas2010}ATLAS Collaboration (G. Aad et al.), ``Charged-particle
multiplicities in pp interactions measured with the ATLAS detector
at the LHC,'' New Journal of Physics \textbf{13} 053033, 1-68 (2011).

\bibitem{pythia6}Torbjörn Sjöstrand, Stephen Mrenna and Peter Skands,
``PYTHIA 6.4 physics and manual,'' Journal of High Energy Physics
\textbf{2006} (05), 026 (2006).

\bibitem{webplotdigitizer}Ankit Rohatgi, WebPlotDigitizer, \url{http://arohatgi.info/WebPlotDigitizer/}

\bibitem{root}Rene Brun and Fons Rademakers, ``ROOT - An Object
Oriented Data Analysis Framework,'' Proceedings AIHENP'96 Workshop,
Lausanne, Sep. 1996, Nucl. Inst. \& Meth. in Phys. Res. \textbf{A
389,} 81-86 (1997). See also: \url{http://root.cern.ch/}

\bibitem{APSsynopsis}Robert Garisto, ``Synopsis: Background checking
at LHC,'' July 6, 2010, \url{http://physics.aps.org/synopsis-for/10.1103/PhysRevLett.105.022002}\end{thebibliography}
\end{document}